\documentclass[10pt]{iopart}
\usepackage{graphicx}
\begin{document}

\title{The effect of dynamical compressive and shear strain on magnetic anisotropy in low symmetry ferromagnetic film}

\author{T. L. Linnik,$^1$ V. N. Kats,$^2$ J. J\"{a}ger,$^3$ A. S. Salasyuk,$^2$ \\D. R. Yakovlev,$^{2,3}$ A. W. Rushforth,$^4$ A. V. Akimov,$^4$ \\ A. M. Kalashnikova,$^2$ M. Bayer,$^{2,3}$ and A. V. Scherbakov$^2$}

\address{$^1$ Department of Theoretical Physics, V. E. Lashkaryov Institute of Semiconductor Physics, 03028 Kyiv, Ukraine\\
$^2$ Ioffe Institute, Russian Academy of Sciences, 194021 St. Petersburg, Russia\\
$^3$ Experimentelle Physik 2, Technische Universit\"{a}t Dortmund, 44221 Dortmund, Germany\\
$^4$ School of Physics and Astronomy, University of Nottingham, Nottingham NG7 2RD, United Kingdom}
\ead{scherbakov@mail.ioffe.ru}
\vspace{10pt}
\begin{indented}
\item[]March 6, 2017
\end{indented}

\begin{abstract}
Dynamical strain generated upon excitation of a metallic film by a femtosecond laser pulse may become a versatile tool enabling control of magnetic state of thin films and nanostructures via inverse magnetostriction on a picosecond time scale. Here we explore two alternative approaches to manipulate magnetocrystalline anisotropy and excite magnetization precession in a low-symmetry film of a magnetic metallic alloy galfenol (Fe,Ga) either by injecting picosecond strain pulse into it from a substrate or by generating dynamical strain of complex temporal profile in the film directly. In the former case we realize ultrafast excitation of magnetization dynamics solely by strain pulses. In the latter case optically-generated strain emerged abruptly in the film modifies its magnetocrystalline anisotropy, competing with heat-induced change of anisotropy parameters. We demonstrate that the optically-generated strain remains efficient for launching magnetization precession, when the heat-induced changes of anisotropy parameters do not trigger the precession anymore. We emphasize that in both approaches the ultrafast change of magnetic anisotropy mediating the precession excitation relies on mixed, compressive and shear, character of the dynamical strain, which emerges due to low-symmetry of the metallic film under study.
\end{abstract}

%
\vspace{2pc}
%
%
%
\ioptwocol

\section{Introduction}\label{Sec:intro}

The lattice symmetry sets up the common and important feature of all crystalline magnetically-ordered materials: the magnetocrystalline anisotropy (MCA). The MCA determines such parameters of a magnetic medium as magnetization direction, magnetic resonances frequencies, coercive fields etc. Since the MCA relates directly to the lattice, applying stress to a magnetic medium allows modifying static and dynamical magnetic properties of the latter. This effect known as inverse magnetostriction or Villary effect was discovered at the end of 19th century and is widely used in both fundamental research and applications. Inverse magnetostriction plays a tremendous role in scaling magnetic devices down, e.g. in tailoring the MCA parameters of nanometer ferromagnetic films by properly chosen lattice mismatch with substrate and providing sensing mechanism in microelectromechanical systems (MEMS). Recently, the importance of inverse magnetostriction was also recognized in the emerging field of ultrafast magnetism focused at manipulating magnetic state of matter on the (sub-) picosecond time scale \cite{Kirilyuk-RMP2010}.

Magnetization control via the inverse magnetostriction at ultrashort timescale is based on the techniques of generating picosecond strain pulses in solids developed in picosecond ultrasonics \cite{Thomsen-PRL1984}. When an opaque medium is subjected to a pico- or femtosecond laser pulse, the light absorption and the following rapid increase of lattice temperature induces thermal stress in a surface region. This results in generation of a picosecond strain pulse with spatial size down to 10\,nm and broad acoustic spectrum (up to 100 GHz), which propagates from excited surface as coherent acoustic wavepacket. It has been demonstrated experimentally, that injection of such a strain pulse into a thin film of ferromagnet can modify the MCA and trigger the precessional motion of magnetization \cite{Scherbakov-PRL2010}. This experiment has initiated intense experimental and theoretical research activities \cite{Scherbakov-PRL2010,Bombeck-PRB2012,Kim-PRL2012,Jager-APL2013,Yahagi-PRB2014,Afanasiev-PRL2014,Jager-PRB2015,Janusonis-APL2015,Janusonis-SciRep2016} in what is now referred to as ultrafast magnetoacoustics.

High interest to ultrafast magnetoacoustics is driven by a number of features, which are specific to interaction between coherent acoustic excitation and magnetization, and do not occur when other ultrafast stimuli are employed. The wide range of generated acoustic frequencies overlaps with the range of magnetic resonances in the magnetically-ordered media. Furthermore, thin films and nanostructures possess specific magnetic and acoustic modes, and matching their frequencies and wavevectors may drastically increase their coupling efficiency \cite{Bombeck-PRB2012}. Finally, there is a well-developed theoretical and computational apparatus for high-precision modeling of spatial-temporal evolution of the strain pulse and respective modulation of the MCA \cite{Linnik-PRB2011,Tas-PRB1994,Wright-IEEE1995}. These advantages, however, may be exploited only if the strain-induced effects are not obscured by other processes triggered by direct ultrafast laser excitation.

Generally, there are two main approaches to single-out the strain-induced impact on magnetization. The first one is the spatial separation, when the response of magnetization to the strain pulses is monitored at the sample surface opposite to the one excited by a laser pulse. It has been used in a number of experiments with various ferromagnetic materials \cite{Scherbakov-PRL2010,Bombeck-PRB2012,Kim-PRL2012,Jager-APL2013}. The irrefutable advantage of such an approach is that the laser-induced heating of a magnetic medium is eliminated due to spatial separation of the laser-impact area and the magnetic specimen. An alternative approach employs the spectral selection instead, when the initially generated strain with broad spectrum is converted into monochromatic acoustic excitation. In this case the efficiency of its interaction with ferromagnetic material is controlled by external magnetic field, which shifts the magnetic resonance frequency. This approach was realized in the experiments with ferromagnetic layer embedded into acoustic Fabry-Perot resonator \cite{Jager-PRB2015} and by means of lateral patterning of ferromagnetic film or optical excitation resulting in excitation of surface acoustic waves \cite{Yahagi-PRB2014,Janusonis-APL2015,Janusonis-SciRep2016}.

Very recently we have demonstrated that the strain-induced impact on the MCA can be reliably traced even in a ferromagnetic film excited directly by a femtosecond laser pulse, despite the complexity of the laser-induced electronic, lattice and spin dynamics emerging in this case \cite{Kats-PRB2016}. Here we present overview of our recent experimental and theoretical studies of the ultrafast strain-induced effects in ferromagnetic galfenol films, where the dynamical strain serves as a versatile tool to control MCA. Magnetization precession serves in this experiments as the macroscopic manifestation of ultrafast changes of MCA. We demonstrate the modulation of the MCA and the corresponding response of magnetization under two different experimental approaches, when the \textit{strain pulses} are injected into the film from the substrate, and when the strain with a \textit{step-like temporal profile} is optically generated directly in a ferromagnetic film. In the case of direct optical excitation we also compare the strain-induced change of MCA to the conventional change of anisotropy via optically-induced heating emerging, and demonstrate that these two contributions can be unambiguously distinguished and suggest the regimes at which either of them dominates. In these studies we have utilized the specific MCA of low-symmetry magnetostrictive galfenol film grown on a (311)-GaAs substrate, which enables generation of dynamical strain of mixed, compressive and shear character, as compared to the pure compressive strain in high-symmetry structures.

The paper is organized as follows. In Sec.\,\ref{Sec:exp} we describe the sample under study and three experimental geometries which enable us to investigate ultrafast changes of magnetic anisotropy. In Sec.\,\ref{Sec:theoryMag} we describe phenomenologically magnetocrystalline anisotropy of the (311) galfenol film and consider how it can be altered on an ultrafast timescale. The following Sec.\ref{Sec:theoryAc} is devoted to generation of dynamical strain in metallic films of low symmetry. In Secs.\,\ref{Sec:expAc},\ref{Sec:extOpt} we present experimental results and analysis of the magnetization precession triggered by purely acoustical pump and by direct optical excitation and demonstrate that even in the latter case optically-generated strain may be a dominant impact allowing ultrafast manipulation of the MCA.

\section{Experimental}\label{Sec:exp}

\subsection{Sample}

Film of a galfenol alloy Fe$_{0.81}$Ga$_{0.19}$ (thickness $d_\mathrm{FeGa}$=100\,nm) was grown on the (311)-oriented GaAs substrate ($d_\mathrm{GaAs}$=100\,$\mu$m) (Fig.\,\ref{Fig:Exp}(a)). As was shown in our previous works \cite{Scherbakov-PRL2010,Jager-APL2013}, the magnetic film of this content and thickness of 100\,nm facilitates a strong response of the magnetization to picosecond strain pulses. The film was deposited by DC magnetron sputtering at a power of 22\,W in an Ar pressure of 1.6\,mTorr. The GaAs substrate was first prepared by etching in dilute hydrochloric acid before baking at 773\,K in vacuum. The substrate was cooled down to 298 K prior to deposition. Detailed x-ray diffraction studies \cite{Bowe-thesis} revealed that the film is polycrystalline, and the misorientation of crystallographic axes of crystallites, average size of which was of a few nanometers, was not exceeding a few degrees. Therefore, the studied film can be treated as the single crystalline one. The equilibrium value of the saturation magnetization is $M_s$=1.59\,T \cite{Restorff-JAP2012}. The SQUID measurements confirmed that the easy magnetization axis is oriented in the film plane along the [0$\bar{1}$1] crystallographic direction ($y$-axis). In our experiments external DC magnetic field \textbf{B} was applied in the sample plane along the magnetization hard axis, which lies along [$\bar{2}$33] crystallographic direction ($x$-axis). In this geometry magnetization \textbf{M} orients along the applied field if the strength of the latter exceeds $B$=150\,mT. At lower field strengths magnetization is along an intermediate direction between the $x$- and $y$-axes. 

\subsection{Experimental techniques}

\begin{figure}
\includegraphics[width=8.6cm]{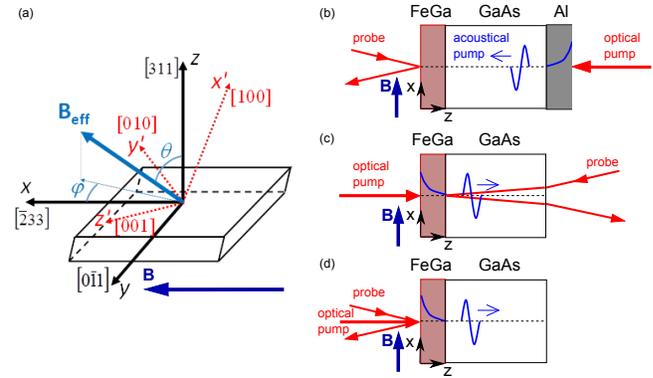}
\caption{(Color online) (a) Schematic presentation of the galfenol film grown on the (311) GaAs substrate. $x'$-, $y'$- and $z'$-axes are directed along the crystallographic [100], [010] and [001] axes, respectively. DC magnetic field \textbf{B} is applied along the [$\bar{2}$33] crystallographic direction, which is the hard magnetization axis. (b-d) Experimental geometries. (b) The optical pump pulses excite 100\,nm thick Al film on the back of the GaAs substrate, thus generating strain pulses injected into the substrate. They act as the acoustical pump triggering the magnetization precession in the galfenol film. The precession is detected by monitoring the rotation of polarization plane of the probe pulses reflected from the galfenol film. (c) The optical pump pulses excite the galfenol film directly. The propagating strain pulses are detected by monitoring polarization rotation for the probe pulses, which penetrate into the GaAs substrate. (d) The optical pump pulses excite the galfenol film directly, increasing the lattice temperature and generating the dynamical strain in the film. Excited magnetization precession is detected by monitoring the rotation of polarization of the probe pulses reflected from the galfenol film. Experiment (b) was performed at $T=$20\,K, experiments (c,d) were performed at room temperature.}
\label{Fig:Exp}
\end{figure}

Three experimental geometries were used in order to explore the impact of dynamical strain on the MCA of the galfenol film. First, the experiments were performed with the dynamical strain being the only stimulus acting on the galfenol film (Fig.\,\ref{Fig:Exp}(b)). A 100-nm thick Al film was deposited on the back side of the GaAs substrate and was utilized as an optoacoustic transducer to inject picosecond strain pulses into the substrate \cite{Thomsen-PRB1986}. The 100-fs optical pump pulses with the central wavelength of 800\,nm, generated by a Ti:sapphire regenerative amplifier, were incident on the Al film inducing rapid increase of its temperature. As a result, as discussed in detail in Sec.\,\ref{Sec:theoryAc}, the picosecond strain pulses were injected into the GaAs substrate. These pulses propagated through the substrate, reached the film (Fe,Ga) film, modified its MCA and triggered the magnetization precession.

The probe pulses split from the same beam were incident on the (Fe,Ga) film at the angle close to 0, and the time-resolved polar magneto-optical Kerr effect (TRMOKE) was measured. In this experimental geometry, TRMOKE rotation angle $\beta_K$ is directly proportional to the out-of-plane deviation of magnetization $\Delta M_z$ induced by a pump:
 \begin{equation}
\Delta \beta_K(t)=\left[\sqrt{\varepsilon_0}(\varepsilon_0-1)\right]^{-1}\chi_{xyz}\Delta M_z(t),\label{Eq:TRMOKE}
\end{equation}
where $\varepsilon_0$ is the diagonal dielectric permittivity tensor component of (Fe,Ga) at the probe wavelength, $\chi_{xyz}$ is the magneto-optical susceptibility at the same wavelength, which enters off-diagonal dielectric permittivity component as $i\varepsilon_{xy}=i\chi_{xyz} M_z$ \cite{Zvezdin-book}. By normalizing TRMOKE rotation by the static one at saturation ($\beta_K^s\sim M_s$), one gets the measure of deviation of the magnetization out of the sample plane, $\Delta M_z(t)/M_s=\Delta\beta_K(t)/\beta_K^s$. These experiments were performed at $T$=20\,K. The choice of low temperature in this experiment was dictated by the fact, that this prevents attenuation of higher frequency components of the strain pulses in the GaAs substrate \cite{Chen-PhMag1994}, thus allowing excitation of precession with high frequency in relatively high applied magnetic fields.

Second and the third types of experiments (Fig.\,\ref{Fig:Exp}(c,d)) were conducted in the geometry, where the (Fe,Ga) film was directly excited by the optical pump pulses. In this geometry there were two contribution to the change of MCA: (i) direct modification of the MCA due to heating \cite{Carpene-PRB2010} and (ii) inverse magnetostrictive effects (See Sec.\,\ref{Sec:theoryMag} for details). In these experiments we used 170-fs pump and probe pulses of the 1030-nm wavelength generated by the Yb:KGd(WO$_4$)$_2$ regenerative amplifier. These experiments were performed at room temperature.

In the second geometry the probe pulses were incident onto the back side of the GaAs substrate (Fig.\,\ref{Fig:Exp}(c)). Since the probe pulses wavelength is well below the GaAs absorption edge, it penetrated the substrate and reached the magnetic film. Thus, here we were able to probe optically excited dynamics of the magnetization of the (Fe,Ga) film. Additionally, this experimental geometry enables one to detect strain pulses injected into the substrate from the film with the velocity $s_j$, where $j$ denotes the particular strain pulses polarization. Upon propagation through GaAs these pulses modified its dielectric permittivity via photoelastic effect. The intensity and the polarization of the probe pulses were therefore modified in the oscillating manner \cite{Thomsen-PRB1986}, with the frequency
\begin{equation}
\nu_j=2s_j\sqrt{\varepsilon_0}\lambda_\mathrm{pr}^{-1},\label{Eq:freqAc}
\end{equation}
where $\lambda_\mathrm{pr}$ is the probe wavelength, $\varepsilon_0$ is the dielectric permittivity of GaAs, and the angle of incidence for the probe pulses is taken to be 0. These oscillations are often referred to as Brillouin oscillations. The main purpose of this experiment was to confirm generation of dynamical strain upon excitation of the (Fe,Ga) film by optical pump pulses.

Third type of experiments was performed in the conventional optical pump-probe geometry, when both optical pump and probe pulses were incident directly on the galfenol film (Fig.\,\ref{Fig:Exp}(d)). This is the main experiment in our study, which demonstrates how various contributions to the optically-induced MCA change can be distinguished and separated.

\section{Thermal and strain-induced control of the magnetic anisotropy in (311) galfenol film}\label{Sec:theoryMag}

The magnetic part of the normalized free energy density of the single crystalline galfenol film $F_M=F/M_s$ grown on the (311)-GaAs substrate (Fig.\,\ref{Fig:Exp}(a)) can be expressed as
\begin{eqnarray}
F_M(\mathbf{m})&=&-\mathbf{m}\cdot\mathbf{B}+B_dm^2_z\label{Eq:energy}\\
&+&K_1\left(m^2_{x'}m^2_{y'}+m^2_{z'}m^2_{y'}+m^2_{x'}m^2_{z'}\right)-K_um_y^2\nonumber\\
&+&b_1(\epsilon_{x'x'}m^2_{x'}+\epsilon_{y'y'}m^2_{y'}+\epsilon_{z'z'}m^2_{z'})\nonumber\\
&+&b_2(\epsilon_{x'y'}m_{x'}m_{y'}+\epsilon_{x'z'}m_{x'}m_{z'}+\epsilon_{y'z'}m_{y'}m_{z'}),\nonumber
\end{eqnarray}
where $\mathbf{m}=\mathbf{M}/M_s$. Here for a sake of convenience Zeeman, shape, and uniaxial anisotropy terms are written in the coordinate frame associated with the film, i.e the $z$-axis is directed along the sample normal. Cubic anisotropy term and the magneto-elastic terms are written in the frame given by the crystallographic axes $x'y'z'$ (Fig.\,\ref{Fig:Exp}(a)). Strain components $\epsilon_{ij}$ are considered to be zero at equilibrium. Corresponding equilibrium orientation of magnetization is given by the direction of an effective magnetic field expressed as
\begin{equation}
\mathbf{B}_\mathrm{eff}=-\frac{\partial F_M(\mathbf{m})}{\partial\mathbf{m}}.\label{Eq:Beff}
\end{equation}

Rapid change of any of the terms in Eq.\,(\ref{Eq:energy}) under an external stimulus may result in reorientation of the effective field (\ref{Eq:Beff}). This and thus trigger magnetization precession, which can be described by the Landau-Lifshitz equation \cite{LL,Gurevich-book}:
\begin{equation}
\frac{d\mathbf{m}}{d t}=-\gamma\cdot\mathbf{m}\times\mathbf{B}_\mathrm{eff}(t),\label{Eq:LL}
\end{equation}
where $\gamma$ is the gyromagnetic ratio. This precession plays a two-fold role. On the one hand, magnetization precession triggered by an ultrafast stimulus is in itself an important result attracting a lot of attention nowadays. On the other hand, magnetization precession is the macroscopical phenomenon, which can be easily observed in conventional pump-probe experiments and, at the same time, allows getting a insight into the complex microscopical processes triggered by various ultrafast stimuli.

We exclude from the further discussion the ultrafast laser-induced demagnetization \cite{Beaurepaire-PRL1996}, which may trigger the magnetization precession \cite{Koopmans-PRL2002} due to the decrease of the demagnetizing field $\mu_0M_s/2$. This contribution to the change of the effective field orientation is proportional to $z$-component of $\mathbf{M}$ at equilibrium, which is zero in the experimental geometry discussed here, with magnetic field applied in the film plane. Thus, we focus on the effects related to the change of the MCA solely and consider two mechanisms.

First mechanism allowing ultrafast change of the MCA relies on \textit{heat-induced} changes of the parameters $K_1$ and $K_u$ in Eq.\,(\ref{Eq:energy}). This phenomenon is inherent to various magnetic metals \cite{Carpene-PRB2010}, semiconductors \cite{Hashimoto-PRL2008}, and dielectrics \cite{deJong-PRB2011}. In metallic films absorption of laser pulse results in subpicosecond increase of electronic temperature $T_e$. Subsequent thermalization between electrons and lattice takes place on a time scale of several picoseconds and yields an increase of the lattice temperature $T_l$. Magnetocrystalline anisotropy of a metallic film, and galfenol in particular, is temperature-dependent \cite{Clark-JAP2005}. Therefore, laser-induced lattice heating results in decrease of MCA parameters. Importantly, this mechanism is expected to be efficient if magnetization is not aligned along the magnetic field \cite{Ma-JAP2015,Shelukhin-arxiv2015}. This can be realized by applying magnetic field of moderate strength along the hard magnetization axis. Otherwise, decrease of $K_{1,u}$ would not tilt $\mathbf{B}_\mathrm{eff}$ already aligned along $\mathbf{B}$.

Second mechanism relies on inverse magnetostriction. As it follows from Eq.\,\ref{Eq:energy}, dynamical strain $\hat\epsilon$ induced in a magnetic film can effectively change the MCA. Such dynamical strain can be created in a film either upon injection from the substrate \cite{Scherbakov-PRL2010}, or due to the thermal stress induced by rapid increase of the lattice temperature by optical pulse. It is important to emphasise, that, in contrast to the heat-induced change of the magnetocrystalline anisotropy constants, the \textit{strain-induced} mechanism can be efficient even if the $\mathbf{B}_\mathrm{eff}$ is aligned along $\mathbf{B}$, if the symmetry of the film and the polarization of the dynamical strain are properly chosen.

\section{Optical generation of the dynamical compressive and shear strain in a metallic film on a low-symmetry substrate}\label{Sec:theoryAc}

\subsection{Optical generation of the dynamical strain}

Increase of the electronic $T_e$ and lattice $T_l$ of a metallic film excited by a femtosecond laser pulse is described by the coupled differential equations:
\begin{eqnarray}
C_e\frac{\partial T_e}{\partial t}&=&\kappa\frac{\partial^2T_e}{\partial z^2}-G(T_e-T_l)+P(z,t);\nonumber\\
C_l\frac{\partial T_l}{\partial t}&=&-G(T_l-T_e),\label{Eq:TwoTemp}
\end{eqnarray}
where $P(z,t)=I(t)(1-R)\alpha\exp{(-\alpha z)}$ is the absorbed optical pump pulse power density, with $I(t)$ describing the Gaussian temporal profile, $\alpha$ is the absorption coefficient, $R$ is the reflection Fresnel coefficient. $C_e=A_eT_e$ and $C_l$ are the specific electronic and lattice heat capacities, respectively; $\kappa$ - the thermal conductivity, $G$ - the electron-phonon coupling constant, considered to be temperature independent. $T_l$ stands for the lattice temperature. Heat conduction to the substrate is usually much less than the one within the film and, thus, is neglected. The boundary conditions are $\partial T_e/\partial z=0$ at $z$=0, and $T_e$=$T_l$=$T$ at $z=\infty$, where $T$ is the initial temperature.

Lattice temperature increase sets up the thermal stress, which in turn leads to generation of dynamical strain \cite{Thomsen-PRB1986,Tas-PRB1994,Wright-IEEE1995}. Details of this process are determined by the properties of the metallic film and of the interface between the metallic film and the substrate. As a generalization, we consider the strain mode with the polarization vector $\mathbf{e}_j$ and the amplitude $u_{0,j}$. Following the procedure described in \cite{Wright-IEEE1995} for a high-symmetry film we express the displacement amplitude in the frequency domain as
\begin{eqnarray}
\delta T_e(z,\omega)&=&\frac{\alpha(1-R)}{\kappa}\frac{I(\omega)}{\alpha^2-p_T^2}\left[\e^{-\alpha z}+\frac{\alpha}{p_T}e^{-p_Tz}\right];\label{Eq:tempE}\\
\delta T_l(z,\omega)&=&\frac{\delta T_e(z,\omega)}{1-i\omega C_lG^{-1}};\label{Eq:tempL}\\
u_{0,j}(z,\omega)&=&\sigma_j\left(\frac{e^{-\alpha z}}{\alpha^2+k_j^2}-\frac{e^{-p_Tz}}{p_T^2+k_j^2}\right.\nonumber\\
&+&\frac{e^{k_jz}}{2ik_j}\left[\frac{1}{\alpha+ik_j}-\frac{1}{p_T+ik_j}\right]\nonumber\\
&+&\left.\frac{e^{-k_jz}}{2ik_j}\left[\frac{1}{\alpha-ik_j}+\frac{1}{p_T-ik_j}\right]\right)\nonumber\\
&+&A_je^{ik_jz}+B_je^{-ik_jz},\label{Eq:strain}
\end{eqnarray}
where we introduced the parameters
\begin{eqnarray}
\sigma_j&=&\frac{e_{j,z}}{\rho s_j^2}\frac{\beta C_l}{1-i\omega C_lG^{-1}}\frac{\alpha^2(1-R)I(\omega)}{\kappa(\alpha^2-p_T^2)},\nonumber\\
p_T&=&\sqrt{\frac{-i\omega C_e}{\kappa}\left(1+\frac{C_lC_e^{-1}}{1-i\omega C_lG^{-1}}\right)},\\
k_j&=&\omega s_j^{-1},\,\mathrm{Re}(p_T)>0.\nonumber
\end{eqnarray}
Here $\beta$ is Gruneisen parameter, $\rho$ is the galfenol density. The constants $A_j$ and $B_j$ are determined from the boundary condition at the free surface $z=0$ and at the (311)-(FeGa)/GaAs interface.

From Eq.\,(\ref{Eq:strain}) it can be seen that thermal stress induces two contributions to the strain in the metallic film. First one is maximal at the film surface and decays exponentially along $z$, which is shown schematically in Figs.\,\ref{Fig:Exp}(b-d). In fact, it closely follows spatial evolution of the lattice temperature $T_l$ in Eq.\,(\ref{Eq:tempL}). In the time domain, this contribution emerges on a picosecond time scale following lattice temperature increase and decays slowly towards equilibrium due to the heat transfer to the substrate. Therefore, on the typical time scale of experiment on ultrafast change of the MCA, i.e $\sim$1\,ns, this contribution can be considered as the \textit{step-like strain emergence}. Second contribution describes the picosecond \textit{strain pulse} propagating away from the film surface along $z$ \cite{Scherbakov-OptExp2013}.

\subsection{Injection of compressive and shear dynamical strain pulses into (311)-galfenol film}

First we consider the scenario illustrated in Fig.\,\ref{Fig:Exp}(b). The Al film serving as the optoacoustic transducer, is polycrystalline and, thus, acoustically isotropic. Thus, longitudinal (LA) strain is generated due to optically-induced thermal stress. Its polarization vector is $\mathbf{e}_\mathrm{LA}=(0,0,1)$ and the amplitude is $u_{0,\mathrm{LA}}$. Corresponding strain component is $\epsilon^\mathrm{LA}_{zz}=e_{\mathrm{LA},z}\partial u_{0,\mathrm{LA}}/\partial z$. This strain is purely compressive/tensile. Due to mode conversion at the interface shear strain may be also generated, but the efficiency of this process is low \cite{Hurley-UltraS2000}. After transmission of the strain pulse through the interface between elastically isotropic Al film and anisotropic low-symmetry single crystalline (311)-GaAs substrate, two strain pulses emerge, quasi-longitudinal (QLA) and quasi-transversal (QTA), with the polarization vectors $\mathbf{e}_\mathrm{QLA}$=(0.165,\,0,\,0.986) and $\mathbf{e}_\mathrm{QTA}$=(0.986,\,0,\,-0.165), propagating further to the substrate \cite{Scherbakov-OptExp2013}. Importantly, both QLA and QTA strain pulses have significant shear components. Expressions for the corresponding amplitudes are found in \cite{Scherbakov-OptExp2013} by taking into account interference between LA and TA modes within the film and multiple reflections and mode conversion at the interface. QLA and QTA pulses injected thus into GaAs substrate propagate with their respective sound velocities.

Upon reaching magnetic (Fe,Ga) film these strain pulses can trigger the magnetization precession \cite{Scherbakov-PRL2010,Jager-APL2013}, by modifying magneto-elastic terms in Eq.\,(\ref{Eq:energy}). Since the QTA and QLA pulse velocities in the 100\,$\mu$m (311)-GaAs substrate are $s_\mathrm{QTA}=$2.9\,km$\cdot$s$^{-1}$ and $s_\mathrm{QLA}=$5.1\,km$\cdot$s$^{-1}$ \cite{Popovic-PRB1993,Scherbakov-OptExp2013}, they reach (Fe,Ga) film after 35 and 20\,ns, respectively, and thus, their impact on the magnetic film can be separated in time. Strictly speaking, polarization vectors of QL(T)A in (Fe,Ga) and in GaAs differ, and transformation of the strain pulses upon crossing GaAs/(Fe,Ga) interface should be taken into account. However, since the mismatch is rather small and both QL(T)A strain pulses remain polarized in the $xz$ plane, we neglect it in the analysis. Therefore, in the experimental geometry, shown in Fig.\,\ref{Fig:Exp}(b), propagating strain pulses (\ref{Eq:strain}) are employed to control magnetization.

\subsection{Generation of compressive and shear dynamical strain pulses in (311)-galfenol film}

By contrast to polycrystalline Al film, in the single crystalline (Fe,Ga) film on the (311)-GaAs substrate the elastic anisotropy plays essential role already at the stage of the strain generation \cite{Matsuda-PRL2004}. Two strain components $\epsilon_{xz}$ and $\epsilon_{zz}$ arise due to coupling of thermal stress to QLA and QTA acoustic waves. Their polarizations are $\mathbf{e}_\mathrm{QLA}$=(0.286,\,0,\,0.958) and $\mathbf{e}_\mathrm{QTA}$=(0.958,\,0,\,-0.286) \cite{Kats-PRB2016} in the film coordinate frame $xyz$ (Fig.\,\ref{Fig:Exp}(a)). Corresponding strain components can be found as
\begin{eqnarray}
\epsilon^\mathrm{QL(T)A}_{xz}&=&0.5e_{QL(T)A,x}\frac{\partial u_{0,\mathrm{QL(T)A}}}{\partial z}\nonumber\\
\epsilon^\mathrm{QL(T)A}_{zz}&=&e_{QL(T)A,z}\frac{\partial u_{0,\mathrm{QL(T)A}}}{\partial z},\label{Eq:strainAmpl}
\end{eqnarray}
i.e. the generated strain is of mixed, compressive and shear, character. Both step-like emergence of the strain and propagating strain pulses can modify MCA. Possible contribution from this step-like emergence of the strain to the change of MCA was pointed out in \cite{Zhao-APL2005}, however, no detailed consideration was performed allowing to confirm feasibility of this process. Importantly, since the step-like emergence of the strain closely follows temporal and spatial evolution of the lattice temperature, distinguishing their effect on the magnetic anisotropy can be ambiguous. We note that in the case of optically excited (Fe,Ga) film the QLA and QTA strain pulses will be also injected into GaAs, and can be detected employing the scheme shown in Fig.\,\ref{Fig:Exp}(c).

\section{Magnetization dynamics in the (311) galfenol film induced by picosecond strain pulses}\label{Sec:expAc}

\begin{figure}
\includegraphics[width=8.6cm]{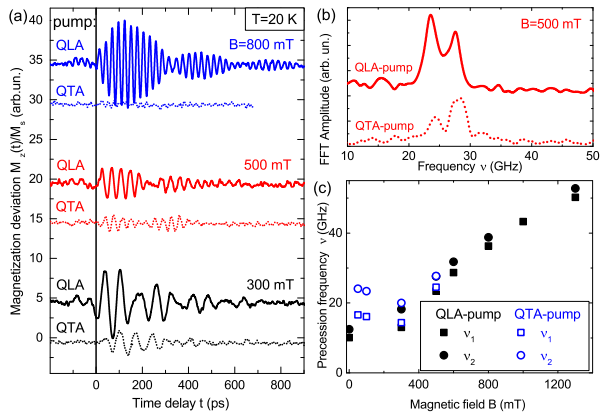}
\caption{(Color online) (a) Probe polarization rotation vs. time delay $t$ measured in the geometry shown in Fig.\,\ref{Fig:Exp}(b) at various values of the magnetic field. $t$=0 is the moment of arrival of the QLA or QTA pulse to the galfenol film, and corresponds to 20 and 35\,ns after the excitation with the optical pump pulse, respectively. (b) FFT spectra of the time delay dependence measured at $B$=500\,mT. The two lines seen in each spectrum correspond to two spin-wave modes (see text for details). (c) Frequency of the probe polarization oscillations caused magnetization precession excited by QLA (closed symbols) and QTA (open symbols) pulses. Optical pump fluence was of $P$=40\,mJ$\cdot$cm$^{-2}$. Note, that in (a) the curves are shifted along the vertical axis for a sake of clarity.}
\label{Fig:AcExPrecession}
\end{figure}

First we examine excitation of the magnetization precession by dynamical strain only, which is realized in the experimental geometry shown in Fig.\,\ref{Fig:Exp}(b). In Fig.\,\ref{Fig:AcExPrecession}(a) we present changes of the probe polarization rotation measured as a function of pump-probe time delay $t$ after QLA or QTA strain pulse arrives to the galfenol film. Time moment $t$=0 for each shown trace corresponds to the time required for either QLA or QTA pulse to travel through the 100\,$\mu$m thick GaAs substrate, and was verified by monitoring reflectivity change \cite{Jager-APL2013}. As one can see, both the QLA and QTA pulses excite oscillations of the probe polarization. Two lines are clearly seen in the Fast Fourier Transform (FFT) spectra of the time traces (Fig.\,\ref{Fig:AcExPrecession}(b)) separated by few GHz. Frequencies of both lines change with the applied field (Fig.\,\ref{Fig:AcExPrecession}(c)), thus confirming that the observed oscillations of the probe polarization originate from the magnetization precession triggered by QLA and QTA strain pulses. The character of the field dependence of $\nu$ (Fig.\,\ref{Fig:AcExPrecession}(c)) corresponds to the one expected for the geometry, when the external magnetic field is applied along the magnetization hard axis. Presence of two field dependent frequencies in the FFT spectra can be attributed to the excitation of two spin wave modes, which is one of the signatures of the magnetization precession excited by picosecond acoustic pulses. As discussed in details in \cite{Bombeck-PRB2012} excitation of several spin waves is enabled by the broad spectrum of the strain pulses and is governed by the boundary conditions in the thin film.

Both the QLA and QTA pulses contain components $\epsilon_{xz}$ and $\epsilon_{zz}$ (\ref{Eq:strainAmpl}). The QLA (QTA) strain pulse enters the magnetic film and propagates though it with the sound velocity of $s_\mathrm{QLA}$=6.0\,km$\cdot$s$^{-1}$ ($s_\mathrm{QTA}$=2.8\,km$\cdot$s$^{-1}$). Upon propagation it contributes to the change of the magneto-elastic term in the free energy in Eq.\,(\ref{Eq:energy}), modifying the MCA of the film, and causing effective magnetic field $\mathbf{B}_\mathrm{eff}$ to deviate from its equilibrium. As a result, magnetization starts to move away from its equilibrium orientation following complex trajectory \cite{Scherbakov-PRL2010}. QL(T)A strain pulse leaves the film after $2d_\mathrm{FeGa}s^{-1}_\mathrm{QL(T)A}$, i.e. 33 and 70\,ps, respectively, and $\mathbf{B}_\mathrm{eff}$ returns to its equilibrium value, while magnetizations relaxes towards $\mathbf{B}_\mathrm{eff}$ precessionally on the much longer nanosecond time scale.

As seen from Fig.\,\ref{Fig:AcExPrecession}(a), amplitude of the magnetization precession excited by QLA phonons is higher than of that excited by QTA phonons. This is in agreement with experimental and theoretical results on propagation of QLA and QTA phonons through the (311)-GaAs substrate \cite{Scherbakov-OptExp2013}, which showed that the amplitude of the displacement associated with the QTA pulses is smaller by a factor of $\sim$5 than that of QLA, while the magnetoelastic coefficients for the shear and compressive strain are the same in galfenol.

Thus, the experiment on excitation of the (311)-(Fe,Ga) film by the picosecond strain pulses clearly demonstrates that dynamical strain effectively excites the magnetization precession in the film in the fields upto 1.2\,T, i.e. when the equilibrium magnetization is already along the applied magnetic field. We note that here we reported the magnetization excitation in the particular geometry, when the magnetic field is applied along the magnetization hard axis. Previously some of the authors also demonstrated analogous excitation in (Fe,Ga) with the field applied in the (311) plane at 45$^\mathrm{o}$ to [$\bar{2}$33] direction, as well as in the field applied along the [311] axis \cite{Jager-APL2013}. It has been also shown that all the features of the excitation observed at low temperature remain valid at room temperature as well. Thus, reported here results obtained at $T$=20\,K can be reliably extrapolated to the room temperature, at which the direct optical excitation of the precession in the galfenol film was studied.

\section{Magnetization dynamics in (311) galfenol film induced by direct optical excitation}\label{Sec:extOpt}

\begin{figure}
\includegraphics[width=8.6cm]{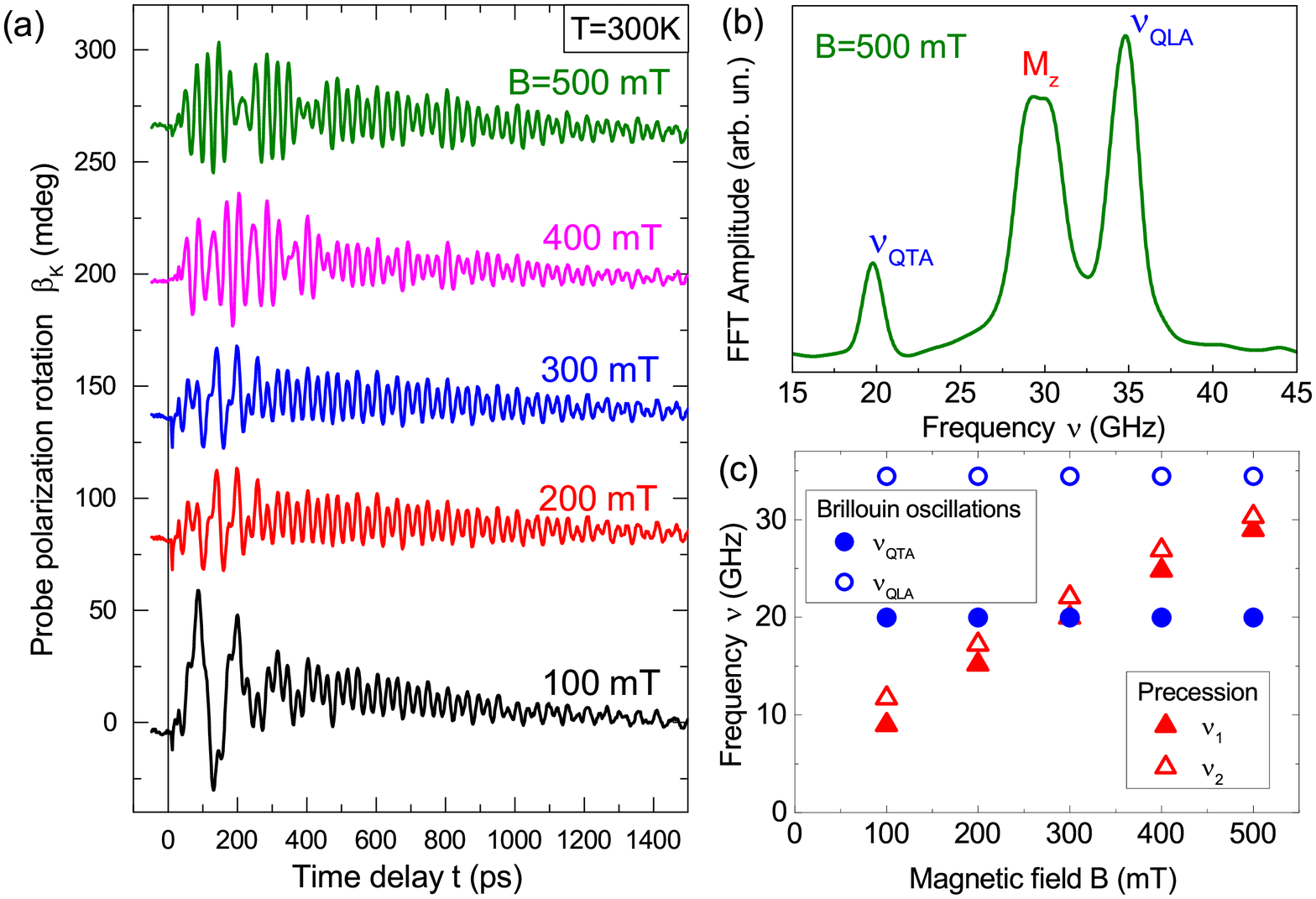}
\caption{(Color online) (a) Probe polarization rotation vs. time delay $t$ measured in the geometry shown in Fig.\,\ref{Fig:Exp}(c) at various magnetic fields. (b) FFT spectrum of the time delay dependence measured at $B$=500\,mT. $\nu_\mathrm{QLA}$, $\nu_\mathrm{QTA}$, and $M_z$ denote the lines corresponding to the Brillouin oscillations related to the QLT and QTA strain pulses, and to the magnetization precession, respectively. (c) Brillouin frequencies in the probe polarization oscillations related to the QTA (closed circles) and QLA (open circles) pulses, and the frequencies related to the optically excited magnetization precession (triangles). Optical pump fluence was of $P$=10\,mJ$\cdot$cm$^{-2}$.}
\label{Fig:DirExAcoustics}
\end{figure}

While in the experiments described in Sec.\,\ref{Sec:expAc} picosecond strain pulses are the only stimulus driving the magnetization precession, the processes triggered by direct optical excitation of a metallic magnetic film are more diverse, and may contribute to both strain-related and other driving forces (see Sec.\,\ref{Sec:theoryMag}). First, in order to confirm generation of dynamical strain in the optically-excited galfenol film we have detected propagating QLA and QTA strain pulses by measuring the polarization rotation for the probe pulses incident onto the back side of the (311)-(Fe,Ga)/GaAs sample (Fig.\,\ref{Fig:Exp}(c)). Fig.\,\ref{Fig:DirExAcoustics}(a) shows the time traces obtained at various magnetic fields. There are several oscillating components clearly present, as can be seen from the Fourier spectra in Fig.\,\ref{Fig:DirExAcoustics}(b). The field dependences of these frequencies are shown in Fig.\,\ref{Fig:DirExAcoustics}(c). The lines at $\nu_\mathrm{QTA}$=20\,GHz and $\nu_\mathrm{QLA}$=35\,GHz are field-independent and are attributed to the Brillouin oscillations caused by the QTA and QLA strain pulses (\ref{Eq:freqAc}), respectively, propagating away from the galfenol film towards the back side of the GaAs substrate with the velocities $s_\mathrm{QTA}<s_\mathrm{QLA}$.

A line in the FFT spectra marked in Fig.\,\ref{Fig:DirExAcoustics}(b) as $M_z$ and possessing the field dependent frequency $\nu$ corresponds to the optically triggered precession of the magnetization in the galfenol film.  This experiment, therefore, confirms concomitant generation of the dynamical strain and excitation of the magnetization precession in the optically excited galfenol film. The mechanism behind the precession excitation is, however, more intricate than in the case of injection of strain pulses in the film.

\begin{figure}
\includegraphics[width=8.6cm]{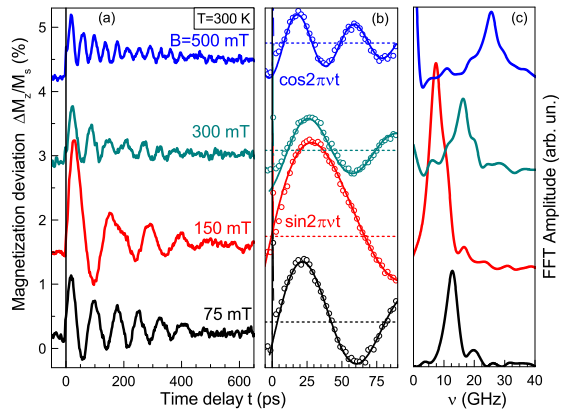}
\caption{(Color online) (a) Probe polarization rotation vs. time delay $t$ measured in the geometry shown in Fig.\,\ref{Fig:Exp}(d) at various values of the magnetic field. (b) The same traces measured with higher temporal resolution. (c) FFT spectra of the time delay dependence. Optical pump fluence was of $P$=10\,mJ$\cdot$cm$^{-2}$.}
\label{Fig:DirExPrecession}
\end{figure}

\begin{figure}
\includegraphics[width=8cm]{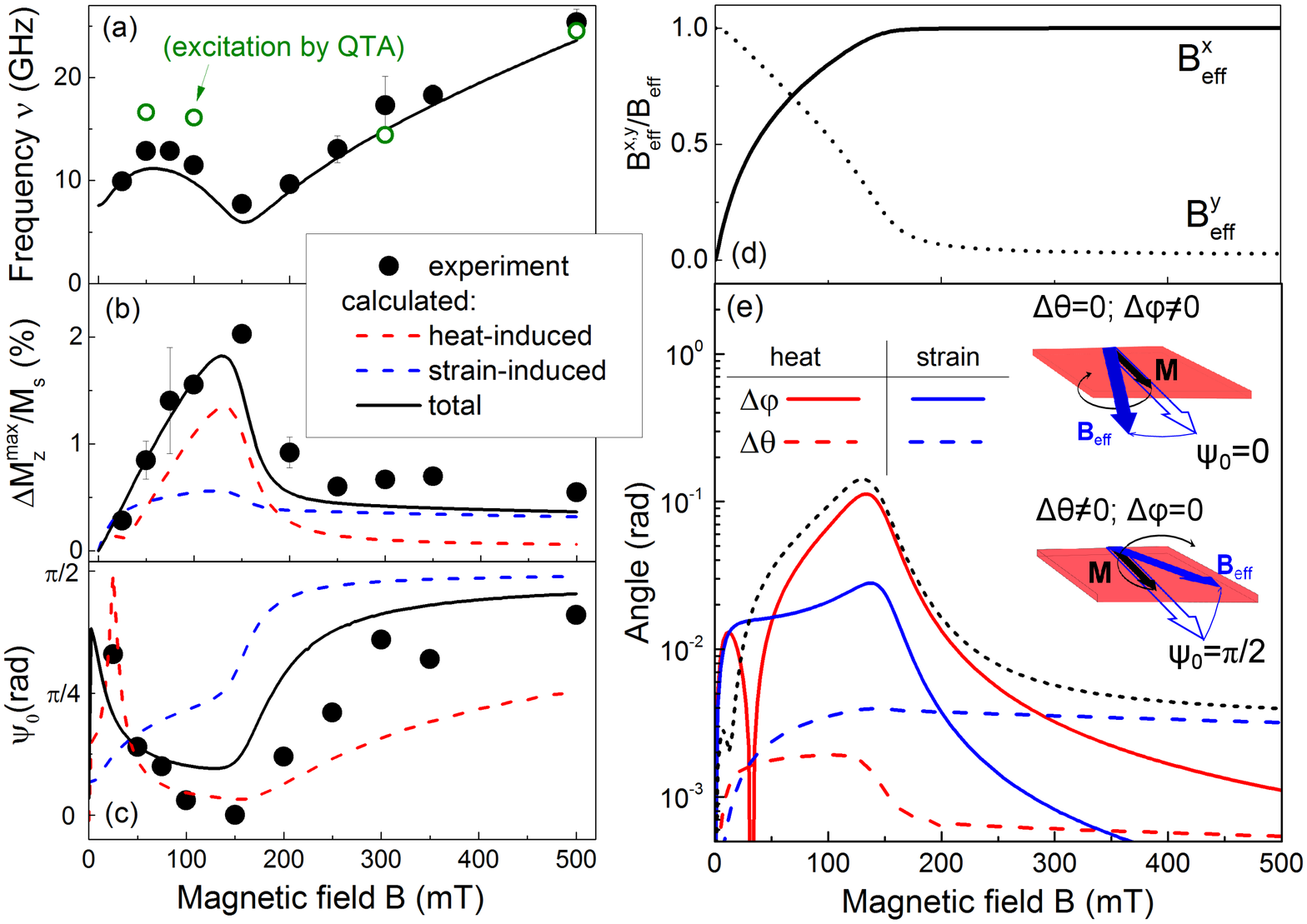}
\caption{(Color online) Field dependencies of (a) the frequency $\nu$, (b) amplitude $M_{z0}/M_s$ and (c) initial phase $\psi_0$ of the magnetization precession excited optically and detected in the geometry shown in Fig.\,\ref{Fig:Exp}(d). Lines show the results of the calculations (see text): solid lines show the results obtained when both heat- and strain-induced contribution to the anisotropy change are taken into account; blue and red dashed lines show the results obtained when only the heat- (red) or strain-induced (blue) change of the MCA is considered. Open symbols in (a) show the frequency of the magnetization excited acoustically by QTA strain pulse (see also Fig.\,\ref{Fig:AcExPrecession}(b). (d) Calculated equilibrium in-plane components of the effective field $\mathbf{B}_\mathrm{eff}$. (e) Field dependences of the $\textbf{B}_\mathrm{eff}$ tilt angles $\Delta\theta$ (dashed lines) and $\Delta\phi$ (solid lines) under the optical excitation resulting in the increase of temperature by 120\,K from equilibrium one (RT). The red and blue lines show the results when only heat-related or strain-related mechanisms is taken into account, respectively. Short-dashed black line shows the net tilt of $\textbf{B}_\mathrm{eff}$, induced by both mechanisms. The upper and lower insets show two cases of purely in-plane and out-of-plane tilts of $\textbf{B}_\mathrm{eff}$ respectively.
}
\label{Fig:DirExFieldDep}
\end{figure}

In order to get an insight into the problem of direct optical excitation of (311)-(Fe,Ga) film we have performed experiment in the geometry shown in Fig.\,\ref{Fig:Exp}(d), with both pump and probe pulses incident directly on the galfenol film. Figs.\,\ref{Fig:DirExPrecession}(a,b) show the temporal evolution of the TRMOKE signal following excitation of the sample by femtosecond laser pulses. FFT spectra (Fig.\,\ref{Fig:DirExPrecession}(c)) contain one line with field dependent frequency (Fig.\,\ref{Fig:DirExFieldDep}(a)). Thus, we observe excitation of the magnetization precession. Oscillatory component in the observed signal can be approximated by the function
\begin{equation}
\frac{\Delta M_z(t)}{M_s}=\frac{\Delta M_z^\mathrm{max}}{M_s}e^{-t/\tau}\sin(2\pi\nu t+\psi_0).\label{Eq:sine}
\end{equation}
As can be seen from Figs.\,\ref{Fig:DirExFieldDep}(a,b) the frequency $\nu$ is minimal and the amplitude $\Delta M_z^\mathrm{max}/M_s$ is maximal at $B$=150\,mT, i.e. when the magnetization becomes parallel to the external field. This is a conventional behaviour, if the magnetic field is applied along the magnetization hard axis. We also note that the frequency vs. applied field dependences in the case of strain-induced and direct optical excitation resemble each other (see Figs.\,\ref{Fig:AcExPrecession}(c) and \ref{Fig:DirExFieldDep}(a)), with some deviation observed at low fields, which could be due to a fact that the measurements were performed at $T$=20 and 293\,K, respectively.

We have studied the evolution of the magnetization right after the direct optical excitation in more detail (Fig.\,\ref{Fig:DirExPrecession}(b)) and, in particular, determined the initial phases $\psi_0$ of precession (\ref{Eq:sine}). The initial evolution of the magnetization suggest that $\mathbf{B}_\mathrm{eff}$ demonstrates a step-like jump from its equilibrium orientation upon the optical excitation and remains in this orientation for the time much longer than the precession decay. This is opposite to the case of injected strain pulse \cite{Scherbakov-PRL2010}, when the magnetization takes a complex path before the harmonic oscillations start, which reflects the fact that $\mathbf{B}_\mathrm{eff}$ follows the strain while it propagates through the film and returns back to equilibrium orientation once the strain pulse has left the film.

The most striking result is that the initial phase $\psi_0$ of the oscillations possesses non-monotonous field dependence. In particular $M_z(t)$ demonstrates pure $sine-$like behaviour when the magnetic field is of $B$=150\,mT, and pure $cosine-$like behavior at $B$=500\,mT. Detailed field dependence of the precession initial phase is shown in Fig.\,\ref{Fig:DirExFieldDep}(c). Keeping in mind that at $t=0$  at any strength of the in-plane magnetic field the magnetization is oriented in the film plane, one concludes that the \textit{sine}-like ($\psi_0=0$) temporal evolution of $M_z$ at the applied field of $B$=150\,mT corresponds to the magnetization precessing around the transient effective field $\mathbf{B}_\mathrm{eff}(t)$, which lies in the sample plane. By contrast, \textit{cosine}-like ($\psi_0=\pi/2$) behavior of the $M_z$ corresponds to the precession around $\mathbf{B}_\mathrm{eff}(t)$, having finite out-of-plane component.

Observed change of the initial phase of the magnetization precession gives a hint that there are, in fact, two competing mechanism of the precession excitation, which relative and absolute efficiencies change with increase of the applied magnetic field. In experiment with optical excitation (Fig.\,\ref{Fig:Exp}(d)) two mechanisms, heat- and strain-induced ones, considered in Sec.\,\ref{Sec:theoryMag} are expected to affect the magnetic anisotropy of the galfenol film. The heat-induced mechanism, based on the rapid increase of the temperature and decrease of the MCA constants, is expected to trigger the precession at relatively low fields. The strain-induced mechanism, resulting from the thermally-induced stress, in turn, can be efficient at high fields as well. The latter is demonstrated in our experiments with purely acoustic excitation of the studied film, where the precession is observed in the applied fields upto at least 1.2\,T (Fig.\,\ref{Fig:AcExPrecession}(a,c))).

In order to test this model we calculated the changes of the MCA parameters $K_{1,u}$ and the magneto-elastic part of the free energy (\ref{Eq:energy}) of the optically-excited (Fe,Ga) film, using the routine described briefly in Secs.\,\ref{Sec:theoryMag} and \ref{Sec:theoryAc} and in more detail in \cite{Kats-PRB2016}. Since some required parameters are unknown for galfenol, for calculations we used those of Fe: $A_e$=672\,J$\cdot$m$^{-3}$K$^{-2}$ \cite{Tari}, $C_l$=3.8$\cdot10^{6}$\,J$\cdot$m$^{-3}$K$^{-1}$, $\kappa$=80.4 W$\cdot$m$^{-1}$K$^{-1}$ \cite{Lide}. The electron-phonon coupling constant $G$=8$\cdot$10$^{17}$ W$\cdot$m$^{-3}$K$^{-1}$ was obtained from \cite{Carpene-PRB2010} where the electron-phonon relaxation time equal to $\sim C_eG^{-1}$ was found to be of 250\,fs. The values of $R$ and $\alpha$ were determined experimentally. The galfenol density $\rho$ for Fe$_{0.81}$Ga$_{0.19}$ is estimated to be of 7.95$\cdot$10$^3$ kg$\cdot$m$^{-3}$. The equilibrium magnetic anisotropy parameters $K_1$=30\,mT, $K_u$=45\,mT and the magneto-elastic coefficients $b_1$=-6\,T, $b_2$=2\,T were found using literature data \cite{Restorff-JAP2012,Parkes-SciRep2013,Atulasimha-SMS2007} as well as from the fit of the field dependence of the precession frequency (Fig.\,\ref{Fig:DirExFieldDep}(a)). Fig.\,\ref{Fig:DirExFieldDep}(d) shows calculated equilibrium in-plane orientation of the effective field $\textbf{B}_\mathrm{eff}$, confirming that it aligns with the external field when the latter exceeds $B$=150\,mT, in agreement with the SQUID data (not shown).

The calculations of the laser-induced magnetization dynamics were performed for the optical excitation density of $P$=10\,mJ$\cdot$cm$^{-2}$. For this laser fluence the lattice temperature increase calculated with Eq.\,(\ref{Eq:tempL}) was found to be of $\Delta T_l$=120\,K. Corresponding change of the MCA parameters was found to be of $\Delta K_1$=-4.75\,mT and $\Delta K_u$=-2.2\,mT, and the persistent components of the compressive and shear dynamical strain were found to be $\epsilon_{zz}=\Delta\epsilon_{zz}$=1.2$\cdot10^{-3}$ and $\epsilon_{xz}=\Delta\epsilon_{xz}$=-4$\cdot10^{-4}$. From these values the optically-triggered out-of-plane $\Delta\theta$ and in-plane $\Delta\phi$ deviations of the effective field $\textbf{B}_\mathbf{eff}$ (Fig.\,\ref{Fig:Exp}(a)) were found, as shown in Fig.\,\ref{Fig:DirExFieldDep}(e). As expected, the heat-induced change of the MCA affects the orientation of the effective field predominantly in the range of the applied fields below and close to $B$=150\,mT. By contrast, the strain-induced deviation of the effective field remains significant even when the applied field is as high as 500\,mT. At lower fields this contribution competes with the heat-induced one. The calculations also confirm that the propagating strain pulses also generated by the optical excitation contribute much weakly to the MCA change.

Importantly, at high fields optically-generated strain results in the out-of-plane deviation $\Delta\theta$ of $\textbf{B}_\mathrm{eff}$. At intermediate and low field the combined effect of the heat- and strain-induced anisotropy change results in both $\Delta\theta$ and $\Delta\phi$ to be non-zero. At $B$=150\,mT, i.e. when equilibrium magnetization is aligned along the external field, the heat-induced change of magnetocrystalline constants dominates and $\textbf{B}_\mathrm{eff}$ deviates mostly in plane. These two limiting situations are illustrated in the insets of Fig.\,\ref{Fig:DirExFieldDep}(e).

Finally, the amplitude and the initial phase of the magnetization precession triggered via heat- and strain-induced change of magnetic anisotropy of galfenol film were calculated (Fig.\ref{Fig:DirExFieldDep}(b,c)). Good agreement between the experimental data and the calculated one confirms that the MCA in the optically-excited (311)-(Fe,Ga) film is indeed modified and, thus, triggers the precession, via two distinct mechanisms. At relatively low fields the heat-induced change of anisotropy parameters is efficient, as was also shown in a number of previous works \cite{Carpene-PRB2010,Shelukhin-arxiv2015}. In addition, optically-generated strain modifies MCA via inverse magnetostriction. Furthermore, as the applied field increases, the heat-induced contribution to the magnetic anisotropy change decreases more rapidly than the strain-induced one. As a result, in the relatively high fields magnetization precession is excited mostly due to the optically-generated persistent strain.

It is instructive to note, that the physical origin of the MCA change - inverse magnetostriction - is the same when either the step-like strain is induced optically in the magnetic film, or the strain pulse is injected into the film. However, the magnetization precession trajectories may appear to be very distinct. In the case of the direct optical excitation the temporal profile of the $\mathbf{B}_\mathrm{eff}$ modified due to abruptly emerged strain can be seen as the step-like jump. This sets the amplitude and initial phase of the magnetization precession, which are uniquely linked to the angle between the equilibrium and modified directions of $\mathbf{B}_\mathrm{eff}$ \cite{Kats-PRB2016}. This situation becomes much more intricate when the strain pulse drives the MCA change. As discussed in \cite{Scherbakov-PRL2010}, strain pulse alternates the direction of $\mathbf{B}_\mathrm{eff}$ upon propagation through the film triggering the magnetization precession, which then proceeds around the equilibrium $\mathbf{B}_\mathrm{eff}$ ones the strain pulse left the film. Thus, the amplitude and the initial phase of the excited precession would be dependent on the particular spatial and temporal profile of the strain pulse.

\section{Conclusions}

In conclusion, we have demonstrated two alternative approaches allowing one to modify magnetocrystalline anisotropy of a metallic magnetic film at ultrafast time scale by dynamical strain, with inverse magnetostriction being the underlying mechanism. Using 100-nm film of a ferromagnetic metallic alloy (Fe,Ga) grown on a low symmetry (311)-GaAs substrate, we were able to trigger the magnetization precession by dynamical strain of a mixed, compressive and shear, character.

Picosecond quasi-longitudinal and quasi-transversal strain pulses can be injected into the galfenol film from the substrate, which leads to efficient excitation of the magnetization precession. In this case, owing to the distinct propagation velocities of QLA and QTA pulses in the substrate, their impact on MCA can be easily distinguished and analyzed. Importantly, dynamical strain remains the efficient stimulus triggering the magnetization precession in the applied magnetic fields up to 1.2\,T.

Alternatively, one can directly excite the galfenol film by a femtosecond laser pulse. In this case there are two competing mechanism mediating the ultrafast change of MCA. Rapid increase of the lattice temperature results in the decrease of the MCA parameters. The lattice temperature increase also sets up the thermal stress which results in optically generated strain. We demonstrate that the heat-induced decrease of the MCA parameters and the change of MCA mediated by the inverse magnetostriction compete and both can trigger the magnetization precession. Despite the fact that the temporal and the spatial profiles of the lattice temperature increase and optically-generated strain closely resemble each other, their impact on MCA can be distinguished. This is possible owing to distinct response of magnetization to the heat-induced \textit{decrease} of anisotropy parameters and to the strain-induced \textit{change and reorientation} of the effective field describing the MCA. In the former case the magnetization precession is triggered only if the external magnetic field applied along the magnetization hard axis is of moderate strength. In the latter case this constrain is lifted and the precession excitation was observed in the applied fields as strong as at least 0.5\,T. The experiments with strain pulses injected into the film suggest that strain-induced precession excitation would remain efficient at higher applied field values as well.

We would like to emphasise, that in order to realize the strain-induced control of magnetic anisotropy in the optically excited metallic film, low symmetry and elastic anisotropy of the latter is of primary importance. As we have shown in our recent study \cite{Kats-PRB2016} in the galfenol film of high symmetry grown on the (001) GaAs substrate magnetic anisotropy change is dominated by the lattice heating and related decrease of MCA parameters. Thus, low-symmetry films and structures, where the dynamical strain can efficiently modify MCA at relatively high magnetic fields, may be promising objects, when one aims at the excitation of magnetization precession of high frequency. This finding highlights further importance of low-symmetry ferromagnetic structures for ultrafast magneto-acoustic studies \cite{Chudnovsky-PRAppl2016}.

Finally, we note that direct detection of the optically-generated quasi-persistent strain in a metallic film is a challenging task. This strain component is intrinsically accompanied by the laser-pulse induced lattice heating. Therefore, detection of this strain component by optical means is naturally obscured by the change of optical properties of the medium \cite{Thomsen-PRB1986}. Realized in our experiments at the high magnetic field limit observation of the magnetization precession around new MCA direction, which was set mostly by the emerged quasi-persistent strain, can be considered as an indirect way to probe this constituent of the ultrafast lattice dynamics.

\section{Acknowledgements}
This work was supported by the Russian Scientific Foundation [grant number 16-12-10485] through funding the experimental studies at the Ioffe Institute; and by the Engineering and Physical Sciences Research Council [grant number EP/H003487/1] through funding the growth and characterization of Galfenol films. The experimental work at TU Dortmund was supported by the Deutsche Forschungsgemeinschaft in the frame of Collaborative Research Center TRR 160 (project B6). The Volkswagen Foundation [Grant number 90418] supported the theoretical work at the Lashkaryov Institute. A.V.A. acknowledges the support from Alexander von Humboldt Foundation.\\


\begin{thebibliography}{99}
\bibitem{Kirilyuk-RMP2010} A. Kirilyuk, A. V. Kimel, and T. Rasing, Ultrafast optical manipulation of magnetic order, Rev. Mod. Phys. \textbf{82}, 2731 (2010).
\bibitem{Thomsen-PRL1984} C. Thomsen, J. Strait, Z. Vardeny, H. J. Maris, J. Tauc, and J. J. Hauser, Coherent phonon generation and detection by picosecond light pulses, Phys. Rev. Lett. \textbf{53}, 989 (1984).
\bibitem{Scherbakov-PRL2010} A. V. Scherbakov, A. S. Salasyuk, A. V. Akimov, X. Liu, M. Bombeck, C. Br\"{u}ggemann, D. R. Yakovlev, V. F. Sapega, J. K. Furdyna, and M. Bayer, Coherent magnetization precession in ferromagnetic (Ga,Mn)As induced by picosecond acoustic pulses, Phys. Rev. Lett. \textbf{105}, 117204 (2010).
\bibitem{Bombeck-PRB2012} M. Bombeck, A. S. Salasyuk, B. A. Glavin, A. V. Scherbakov, C. Br\"{u}ggemann, D. R. Yakovlev, V. F. Sapega, X. Liu, J. K. Furdyna, A. V. Akimov, and M. Bayer, Excitation of spin waves in ferromagnetic (Ga,Mn)As layers by picosecond strain pulses, Phys. Rev. B \textbf{85}, 195324 (2012).
\bibitem{Kim-PRL2012} J.-W. Kim, M. Vomir, and J.-Y. Bigot, Ultrafast magnetoacoustics in nickel films, Phys. Rev. Lett. \textbf{109}, 166601 (2012).
\bibitem{Jager-APL2013} J. V. J\"{a}ger, A. V. Scherbakov, T. L. Linnik, D. R. Yakovlev, M. Wang, P. Wadley, V. Holy, S. A. Cavill, A. V. Akimov, A. W. Rushforth, and M. Bayer, Picosecond inverse magnetostriction in galfenol thin films, Appl. Phys. Lett. \textbf{103}, 032409 (2013).
\bibitem{Yahagi-PRB2014} Y. Yahagi, B. Harteneck, S. Cabrini, and H. Schmidt, Controlling nanomagnet magnetization dynamics via magnetoelastic coupling, Phys. Rev. B \textbf{90}, 140405(R) (2014).
\bibitem{Afanasiev-PRL2014} D. Afanasiev, I. Razdolski, K. M. Skibinsky, D. Bolotin, S. V. Yagupov, M. B. Strugatsky, A. Kirilyuk, Th. Rasing, and A. V. Kimel, Laser excitation of lattice-driven anharmonic magnetization dynamics in dielectric FeBO$_3$, Phys. Rev. Lett. \textbf{112}, 147403 (2014).
\bibitem{Jager-PRB2015} J. V. J\"{a}ger, A. V. Scherbakov, B. A. Glavin, A. S. Salasyuk, R. P. Campion, A. W. Rushforth, D. R. Yakovlev, A. V. Akimov, and M. Bayer, Resonant driving of magnetization precession in a ferromagnetic layer by coherent monochromatic phonons, Phys. Rev. B \textbf{92}, 020404(R) (2015).
\bibitem{Janusonis-APL2015} J. Janu\v{s}onis, C. L. Chang, P. H. M. van Loosdrecht, and R. I. Tobey, Appl. Phys. Lett. \textbf{106}, 181601 (2015).
\bibitem{Janusonis-SciRep2016} J. Janu\v{s}onis, T. Jansma, C. L. Chang, Q. Liu, A. Gatilova, A. M. Lomonosov, V. Shalagatskyi, T. Pezeril, V. V. Temnov, and R. I. Tobey, Transient grating spectroscopy in magnetic thin films: simultaneous detection of elastic and magnetic dynamics, Sci. Rep. \textbf{6}, 29143 (2016).
\bibitem{Linnik-PRB2011} T. L. Linnik, A. V. Scherbakov, D. R. Yakovlev, X. Liu, J. K. Furdyna, and M. Bayer, Theory of magnetization precession induced by a picosecond strain pulse in ferromagnetic semiconductor (Ga,Mn)As, Phys. Rev. B \textbf{84}, 214432 (2011).
\bibitem{Tas-PRB1994} G. Tas and H. J. Maris, Electron diffusion in metals studied by picosecond ultrasonics, Phys. Rev. B \textbf{49}, 15046 (1994).
\bibitem{Wright-IEEE1995} O. B. Wright and V. E. Gusev, Ultrafast generation of acoustic waves in copper, IEEE Trans. Ultrason. Ferroelectr. Freq. Control. \textbf{42}, 331 (1995).
\bibitem{Kats-PRB2016} V. N. Kats, T. L. Linnik, A. S. Salasyuk, A. W. Rushforth, M. Wang, P. Wadley, A. V. Akimov, S. A. Cavill, V. Holy, A. M. Kalashnikova, and A. V. Scherbakov, Ultrafast changes of magnetic anisotropy driven by laser-generated coherent and noncoherent phonons in metallic films, Phys. Rev. B \textbf{93}, 214422 (2016).
\bibitem{Bowe-thesis} S. Bowe, \textit{Magnetisation dynamics in magnetostrictive nanostructures} (PhD thesis, The University of Nottingham).
\bibitem{Restorff-JAP2012} J. B. Restorff, M. Wun-Fogle, K. B. Hathaway, A. E. Clark, T. A. Lograsso, and G. Petculescu, Tetragonal magnetostriction and magnetoelastic coupling in Fe-Al, Fe-Ga, Fe-Ge, Fe-Si, Fe-Ga-Al, and Fe-Ga-Ge alloys, J. Appl. Phys. \textbf{111}, 023905 (2012).
\bibitem{Thomsen-PRB1986} C. Thomsen, H. T. Grahn, H. J. Maris, and J. Tauc, Surface generation and detection of phonons by picosecond light pulses, Phys. Rev. B \textbf{34}, 4129 (1986).
\bibitem{Popovic-PRB1993} Z. V. Popovic, J. Spitzer, T. Ruf, M. Cardona, R. N\"{o}tzel, and K. Ploog, Folded acoustic phonons in GaAs/AlAs corrugated superlattices grown along the [311] direction, Phys. Rev. B \textbf{48}, 1659 (1993).
\bibitem{Scherbakov-OptExp2013} A. V. Scherbakov, M. Bombeck, J. V. J\"{a}ger, A. S. Salasyuk, T. L. Linnik, V. E. Gusev, D. R. Yakovlev, A. V. Akimov, and M. Bayer, Picosecond opto-acoustic interferometry and polarimetry in high-index GaAs, Opt. Exp. \textbf{21}, 16473 (2013).
\bibitem{Zvezdin-book} A. K. Zvezdin and V. A. Kotov, \textit{Modern magnetooptics and magnetooptical materials} (CRC Press, Boca Raton, 1997).
\bibitem{Chen-PhMag1994} W. Chen , H. J. Maris , Z. R. Wasilewski, and S.-I. Tamura, Attenuation and velocity of 56 GHz longitudinal phonons in gallium arsenide from 50 to 300 K, Phil. Mag. B \textbf{70}, 687 (1994).
\bibitem{Carpene-PRB2010} E. Carpene, E. Mancini, D. Dazzi, C. Dallera, E. Puppin, and S. De Silvestri, Ultrafast three-dimensional magnetization precession and magnetic anisotropy of a photoexcited thin film of iron, Phys. Rev. B \textbf{81}, 060415 (2010).
\bibitem{LL} L. D. Landau and E. M. Lifshitz, Theory of the dispersion of magnetic permeability in ferromagnetic bodies, Phys. Z. Sowietunion \textbf{8}, 153 (1935).
\bibitem{Gurevich-book} A. G. Gurevich and G. A. Melkov, \textit{Magnetization Oscillations and Waves} (CRC-Press, Boca Raton, 1996).
\bibitem{Beaurepaire-PRL1996} E. Beaurepaire, J.-C. Merle, A. Daunois, and J.-Y. Bigot, Ultrafast spin dynamics in ferromagnetic nickel, Phys. Rev. Lett. \textbf{76}, 4250 (1996).
\bibitem{Koopmans-PRL2002} M. van Kampen, C. Jozsa, J. T. Kohlhepp, P. LeClair, L. Lagae, W. J. M. de Jonge, and B. Koopmans, All-optical probe of coherent spin waves, Phys. Rev. Lett. \textbf{88}, 227201 (2002).
\bibitem{Clark-JAP2005} A. E. Clark, M. Wun-Fogle, J. B. Restorff, K. W. Dennis, T. A. Lograsso, and R. W. McCallum, Temperature dependence of the magnetic anisotropy and magnetostriction of Fe$_{100−x}$Ga$_x$ ($x=$8.6, 16.6, 28.5), J. Appl. Phys. \textbf{97}, 10M316 (2005).
\bibitem{Hashimoto-PRL2008} Y. Hashimoto, S. Kobayashi, and H. Munekata, Photoinduced precession of magnetization in ferromagnetic (Ga,Mn)As, Phys. Rev. Lett. \textbf{100}, 067202 (2008).
\bibitem{deJong-PRB2011} J. A. de Jong, A. V. Kimel, R. V. Pisarev, A. Kirilyuk, and Th. Rasing, Laser-induced ultrafast spin dynamics in ErFeO$_3$, Phys. Rev. B \textbf{84}, 104421 (2011).
\bibitem{Ma-JAP2015} T. P. Ma, S. F. Zhang, Y. Yang, Z. H. Chen, H. B. Zhao, and Y. Z. Wu, Distinguishing the laser-induced spin precession excitation mechanism in Fe/MgO(001) through field orientation dependent measurements, J. Appl. Phys. \textbf{117}, 013903 (2015).
\bibitem{Shelukhin-arxiv2015} L. A. Shelukhin, V. V. Pavlov, P. A. Usachev, R. V. Pisarev, and A. M. Kalashnikova, Ultrafast laser-induced changes of the magnetic anisotropy in iron garnet films, arxiv:1507.07437.
\bibitem{Hurley-UltraS2000} D. H. Hurley, O. B. Wright, O. Matsuda, V. E. Gusev, and O. V. Kolosov, Laser picosecond acoustics in isotropic and anisotropic materials, Ultrasonics \textbf{38}, 470 (2000).
\bibitem{Matsuda-PRL2004} O. Matsuda, O. B. Wright, D. H. Hurley, V. E. Gusev, and K. Shimizu, Coherent Shear Phonon Generation and Detection with Ultrashort Optical Pulses, Phys. Rev. Lett. \textbf{93}, 095501 (2004).
\bibitem{Zhao-APL2005} H. B. Zhao, D. Talbayev, Q. G. Yang, and G. L\"{u}pkea, A. T. Hanbicki, C. H. Li, O. M. J. van $'$t Erve, G. Kioseoglou, and B. T. Jonker, Ultrafast magnetization dynamics of epitaxial Fe films on AlGaAs (001), Appl. Phys. Lett. \textbf{86}, 152512 (2005).
\bibitem{Tari} A. Tari, \textit{The Specific Heat of Matter at Low Temperatures} (London: Imperial College Press, 2003), p. 71.
\bibitem{Lide} D. R. Lide, CRC Handbook of Chemistry and Physics (CRC Press, Florida, 2003).
\bibitem{Parkes-SciRep2013} D. E. Parkes, L. R. Shelford, P. Wadley, V. Holy, M. Wang, A. T. Hindmarch, G. van der Laan, R. P. Campion, K. W. Edmonds, S. A. Cavill, and A. W. Rushforth, Magnetostrictive thin films for microwave spintronics, Sci. Rep. \textbf{3}, 2220 (2013).
\bibitem{Atulasimha-SMS2007} J. Atulasimha, A. B. Flatau, and E. Summers, Characterization and energy-based model of the magnetomechanical behavior of polycrystalline iron–gallium alloys, Smart Mater. Struct. \textbf{16}, 1265 (2007).
\bibitem{Chudnovsky-PRAppl2016} E. M. Chudnovsky and R. Jaafar, Manipulating the magnetization of a nanomagnet with surface acoustic waves: spin-rotation mechanism, Phys. Rev. Appl. \textbf{5}, 031002 (2016).
\end{thebibliography}
\end{document}